\documentclass[a4paper]{article}
\usepackage{xcolor}
\usepackage{INTERSPEECH2020}

\title{CONTEXTUAL RNN-T FOR OPEN DOMAIN ASR}

\name{Mahaveer Jain, Gil Keren, Jay Mahadeokar, Geoffrey Zweig, Florian Metze, Yatharth Saraf
}
\address{
  Facebook AI, USA
  }
\email{\{jainmahaveer,gilkeren,jaym,gzweig,fmetze,ysaraf\}@fb.com}
\begin{document}

\maketitle
\begin{abstract}
 End-to-end (E2E) systems for automatic speech recognition (ASR), such as RNN Transducer (RNN-T) and  Listen-Attend-Spell (LAS) blend the individual components of a traditional hybrid ASR system - acoustic model, language model, pronunciation model - into a single neural network. While this has some nice advantages, it limits the system to be trained using only paired audio and text. Because of this, E2E models tend to have difficulties with correctly recognizing rare words that are not frequently seen during training, such as entity names. 
 In this paper, we propose modifications to the RNN-T model that allow the model to utilize additional metadata text with the objective of improving performance on these named entity words. We evaluate our approach on an in-house dataset sampled from de-identified public social media videos, which represent an open domain ASR task. By using an attention model and a biasing model to leverage the contextual metadata that accompanies a video, we observe a relative improvement of about 16\% in Word Error Rate on Named Entities (WER-NE) for videos with related metadata.
\end{abstract}

\section{Introduction}

Present day ASR models using Deep Neural Networks (DNN) can be broadly classified into two frameworks: hybrid~\cite{hybrid-ctc} and E2E~\cite{rnnt-graves, las, loc_attn}. A typical hybrid HMM-DNN system consists of three components trained individually: an acoustic model (AM) that estimates the  posterior probabilities of Hidden Markov Model (HMM) states, a language model (LM) that estimates probabilities of word sequences, and a pronunciation  model (PM) to map phonemes to words. These models are optimized independently~\cite{hinton2012deep} and then combined together using a Weighted Finite State Transducer (WFST) ~\cite{mohri2002weighted} for efficient decoding. In an E2E speech recognition model such as the RNN-T \cite{rnnt-graves}, a single neural network learns to map audio to text instead of using the distinct components of the hybrid systems.
While this generally simplifies overall training and inference pipelines for ASR, E2E models tend to have difficulties with correctly recognizing words that do not appear frequently in paired audio-text training data \cite{clas,dc_tom,dc_phoebe}. Since hybrid ASR systems optimize AM, LM and PM components independently, they can address the rare word recognition issue by 1) training the LM component with large amounts of unpaired text data to model occurrences of rare words and 2) representing pronunciation in the PM.

In this work, we propose an attention-based context biasing approach to address the following underlying issues in the setting of an E2E RNN-T based ASR system: 

\begin{itemize}
  \item \textbf{Rare word recognition}: It is common for a vanilla RNN-T ASR system to make mistakes in recognizing words that occur infrequently in the training data. Assuming that \emph{PyTorch} is a rare word in paired ASR training data, a vanilla RNN-T model might produce a hypothesis as \emph{when you look at \textbf{pie towards} itself} whereas the true transcript is \emph{when you look at \textbf{PyTorch} itself}. Although \emph{PyTorch} is a rare word in training data, it's appearance in the video metadata can be used to recognize it correctly.
  \item \textbf{Disambiguation between similar words}: When acoustically similar words appear in similar contexts in the training data, it is difficult for the model to pick the correct word without additional biasing. E.g. the names \emph{Sean} and \emph{Shaun} might appear with similar frequencies in a similar context, but we might want to prefer one of them if it appears also in related text metadata.
\end{itemize}

Entity names often suffer from both these issues. Further, these often carry a high degree of semantic meaning relative to other words in the transcript, so it is important for ASR systems to recognize these correctly. Therefore, we measure the effectiveness of our approach on recognition of entity names.
We aim to address these problems by biasing ASR using additional context from the accompanying text metadata of the video. This metadata is unstructured and potentially irrelevant to the speech being transcribed, so the ASR system needs to learn to selectively use or ignore it.

The rest of the paper is organized as follows: We review prior work around use of contextual words in E2E ASR in Section~\ref{sec:priorwork}. We describe the base RNN-T model in Section~\ref{sec:RNN-T}. In Section~\ref{sec:rnnt_dc}, we propose changes to the RNN-T model to allow it to incorporate unpaired and unstructured text context via attention and biasing mechanism. We show experiment results in Section~\ref{sec:exp} and further analyze the effectiveness of the proposed method in Section~\ref{sec:analysis} by visualizing attention weights towards the metadata. %REMOVED: We conclude in Section~\ref{sec:conclusion}.

\section{Prior Work}
\label{sec:priorwork}
Prior work has leveraged contextual words either by on-the-fly (OTF) rescoring ~\cite{shallowe2e, shallowfusion2018, streaming_rnnt} or as an additional input to the DNN along with the audio. The first approach is generally referred to as Shallow Fusion whereas the latter as Deep Contextualization ~\cite{clas}. Our work falls in the latter category. It is most closely related to Contextual Listen, Attend And Spell (CLAS) ~\cite{clas}, which also used context words from unpaired text to bias an E2E ASR model. The CLAS model was originally evaluated for closed domain ASR tasks like those used for virtual assistants by using entities such as contact names as context words. Further improvements to CLAS were done in ~\cite{dc_phoebe} and \cite{dc_tom} by using representations that leverage phonetic information as well. In this work, different from CLAS, we look at Deep Contextualization in the setting of an RNN-T ASR model, and evaluate our method on an open domain video ASR task using noisy text metadata from videos as context. In a closed domain use case such as making calls through an assistant, there is strong prior information about where entity names can appear in the utterance, whereas in our case the context words may appear anywhere in the conversational speech of the video.  Deep contextualization of RNN-T was explored in \cite{keyword-rnnt} for keyword spotting use case, where the phoneme sequence of the keyword represented as a one-hot vector was used to attend to and recognize the target keyword. An alternate approach for using contextual metadata from videos to improve ASR is explored in \cite{darong-paper}, where lattices produced by a hybrid ASR system are rescored using metadata.

\section{RNN Tranducer}
\label{sec:RNN-T}

\begin{figure}[t]
  \centering
  \includegraphics[width=8cm,height=6cm,keepaspectratio]{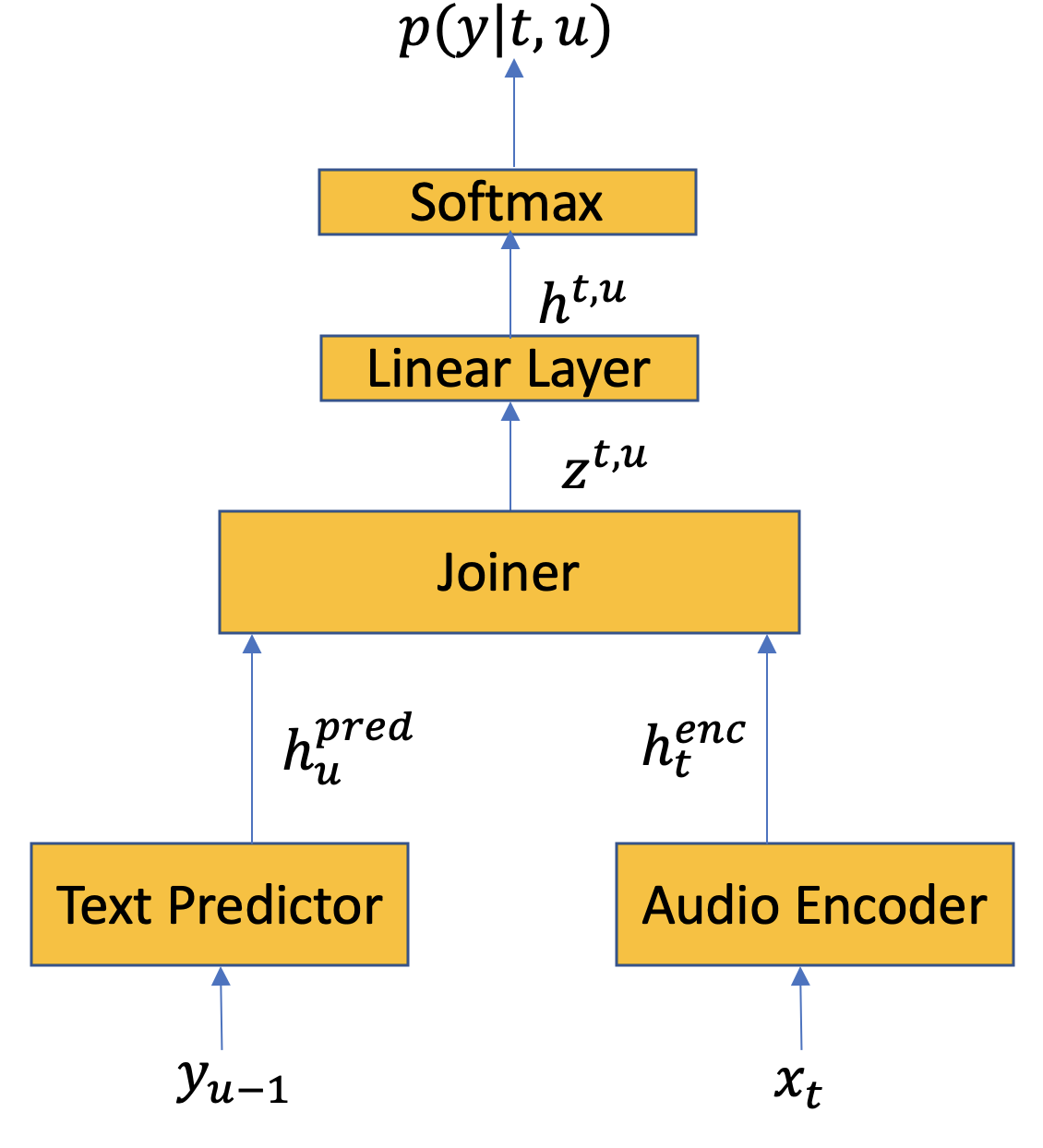}
 \caption{RNN Transducer for ASR}
  \label{fig:rnnt_fig} 
\end{figure}

The framework of RNN-T ASR system is illustrated in Fig.~\ref{fig:rnnt_fig}. RNN-T for ASR has three main components: Audio Encoder, Text Predictor and Joiner. 

The Audio Encoder uses audio frame at $x_{t}$ to produce audio embedding $h^{enc}_{t}$ (Equation~\ref{enc_eq}). The Audio Encoder used in this work is a stack of bi-directional LSTM (BLSTM) layers.
\begin{equation}\label{enc_eq}
	h^{enc}_{t}=f^{enc}(x_{t})
\end{equation}

The Text Predictor uses the last non-blank target unit $y_{u-1}$ to produce embedding $h^{pred}_{u}$ (Equation~\ref{pred_eq}). The Text Predictor is a stack of LSTM layers in this work. We use sentence pieces as target units.

\begin{equation}\label{pred_eq}
	h^{pred}_{u}=f^{pred}(y_{u-1})
\end{equation}

The Joiner takes in the output of Audio Encoder and Text Predictor and combines them to produce an embedding $z^{t, u}$:

\begin{equation}\label{joiner_eq}
    \begin{split}
	z^{t, u} & = \phi(U h^{enc}_{t}+ Vh^{pred}_{u} + b)
	\end{split}
\end{equation}

$U$ and $V$ are matrices that are used to project audio and text embeddings to the same dimensions. $\phi$ is a non-linear function such as Relu~\cite{relu} or tanh.

Finally, the joiner's output, $z^{t, u}$, is passed through a linear transformation followed by a softmax layer to produce  a probability distribution over target units ($y$) , i.e. sentence pieces plus a special $blank$ symbol:
\begin{subequations}

\begin{equation}
\label{softamx_linear_eq}
h^{t, u}  =  W_{y}z^{t, u} + b
\end{equation}

\begin{equation}
\label{softamx_eq}
p(y|t,u) = softmax(h^{t, u})
\end{equation}
\end{subequations}

By incorporating both audio and text for producing $p(y|t,u)$ (Equation~\ref{softamx_eq}), RNN-T can overcome the conditional independence assumption of CTC models~\cite{graves2006connectionist}. The emission of $blank$ as output unit results in an update of the audio embedding by moving ahead in time axis $t$ whereas emission of non $blank$ results in a change in the text embedding. This results in various possible alignment paths as shown in the lattice of size $T*U$ in Figure 1 of \cite{rnnt-graves}. The sum of probabilities of these paths gives the probability of an output sequence, $Y$, given the input sequence, $X$, where $Y$ is the sequence of non $blank$ output target units and $X$ is the input sequence of audio frames.

\section{Contextual RNN-T}
\label{sec:rnnt_dc}

\begin{figure}[t]
  \centering
  \includegraphics[width=8cm,height=12cm,keepaspectratio]{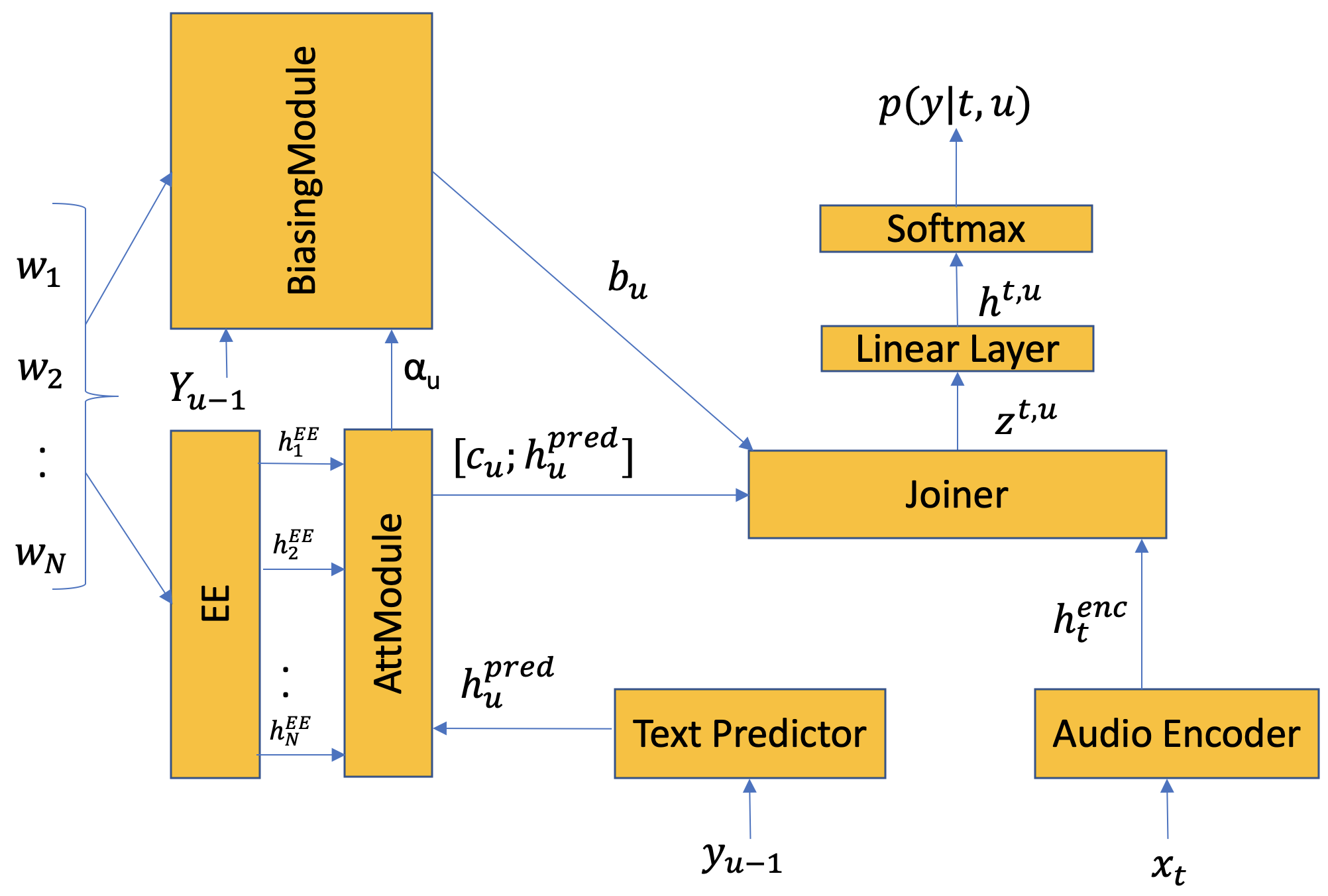}
  %{dc_rnnt_cropped1.png}
 \caption{{Contextual RNN Transducer for ASR}} %Contextualization in Text Predictor of RNN-T}
  \label{fig:rnnt_dc_fig} 
\end{figure}

We modify the base RNN-T model described in Section~\ref{sec:RNN-T} and {\color{black} add three additional components: an Embedding Extractor (EE), an Attention Module (AttModule) and a Biasing Module (BiasingModule)} as shown in Figure \ref{fig:rnnt_dc_fig}.

As in \cite{clas}, each context word, $w_{i}$, is first represented as a sequence of target sentence piece units, e.g. the word ``Jarred" may be mapped to \emph{[Ja, r, re, d]}. This sequence is then fed to an BLSTM, and the last state of the BLSTM is used as the embedding of the given context word (shown as $h^{EE}_{i}$ in Figure \ref{fig:rnnt_dc_fig}).

In the vanilla RNN-T system described in Section \ref{sec:RNN-T}, probabilities over target units  $p(y|t,u)$ (Equation \eqref{softamx_eq}) are conditionally dependent on the outputs of the Audio Encoder, $h^{enc}_{t}$, and Text Predictor, $h^{pred}_{u}$.  In contextual RNN-T, we would like to make $p(y|t,u)$ conditionally dependent on contextual metadata words as well. This dependency can be achieved by incorporating the context word {\color{black}information} %embeddings,  $h^{EE}_{i}$,
into any of the Audio Encoder, Text Predictor and Joiner components. In this work, we explore incorporating the context word
%embeddings,  $h^{EE}_{i}$,
{\color{black}information into the Text Predictor and  Joiner of the RNN-T}. 

An Attention Module (AttModule) is used to compute attention for each word in the metadata text. AttModule uses the predictor output for non-blank text history up to $u$ ($h^{pred}_{u}$) and word embedding, $h^{EE}_{i}$, to compute attention weight, $e_{u,i}$, as shown in Equation \eqref{att_weight_eq}. We use location-aware attention that takes into account the attention weights from the previous predictor state, $\alpha_{u-1}$, while computing alignments at the current step~\cite{loc_attn}. 
\begin{subequations}

\begin{equation}\label{att_conv_eq}
	F = Q * \alpha_{u-1}
\end{equation}

\begin{equation}\label{att_weight_eq}
e_{u,i} = w^\intercal tanh(Ah^{pred}_{u} + Bh^{EE}_{i} + Cf_{i} + b)
\end{equation}
\begin{equation}\label{att_softmax_eq}
\alpha_{u,i} = exp(e_{u,i})/ \sum_{j=1}^{N}(exp(e_{u,j}))
\end{equation}
\end{subequations}

$F$ is output of convolution of $\alpha_{u-1}$ with matrix $Q$ (Equation \eqref{att_conv_eq}).  $f_{i}$ in Equation \eqref{att_weight_eq} is used to denote the output for the $i$th word in $F$. $A$, $B$, $C$ are matrices, $b$ and $w$ are vectors.

We next compute a \emph{context vector}, $c_{u}$, \cite{clas,keyword-rnnt} as the weighted sum of \emph{word embedding} as:  

\begin{equation} \label{context_vector_eq}
c_{u}=\sum_{i=1}^{N}(\alpha_{u,i}*h^{EE}_{i})
\end{equation}. 

We concat this context vector, $c_{u}$, to the output of the Text Predictor ($h^{pred}_{u}$). This results in a modification of the Joiner  equation \eqref{joiner_eq} to equation   \eqref{joiner_eq_modified}:

\begin{equation}\label{joiner_eq_modified}
    \begin{split}
	z^{t, u} & = \phi(U h^{enc}_{t}+ V[c_{u}; h^{pred}_{u}]  + b)
	\end{split}
\end{equation}

{\color{black}An additional biasing module (BiasingModule) may be used to find an active subset of the context words that have the same prefix as the last unfinished word in the text history ($Y_{u-1}$) to create an additional vector that biases the Joiner towards selecting from the subset of sentence pieces that correspond to the active context words using the attention values $\alpha_{u,i}$ from Equation \eqref{att_softmax_eq}. For example, if the decoded form of $Y_{u-1}$ is \textit{Africa An} and the list of the context word is \textit{Android, Antenna} and \textit{Pytorch} then the active context words are \textit{Android} and \textit{Antenna}. The BiasingModule computes a biasing value, $bias_{u,i,k}$, for each sentence piece $k$ of word $w_{i}$ at a given text history $Y_{u-1}$ as shown in Equation \eqref{biasingeq1}. $I_{SP}({w_{i},Y_{u-1}, k})$ returns 1 if $w_{i}$ is active word at $Y_{u-1}$ and $k$ is the sentence piece in $w_{i}$ followed by the shared prefix, otherwise it returns 0. We then compute $bias_{u,k}$ for each sentence piece $k$ at $Y_{u-1}$ by summing over all context words as in Equation \eqref{biasingeq2}. A vector of all biases, $b_{u}$, with length equal to number of sentence pieces is then linearly projected and passed through an optional Dropout layer before feeding to the Joiner as shown in Equation \eqref{joiner_eq_bias_modified} when both BiasingModule and AttModule are used.

\begin{equation}\label{biasingeq1}
bias_{u,i,k} = \alpha_{u,i}*I_{SP}({w_{i},Y_{u-1}}, k)
\end{equation}

\begin{equation}\label{biasingeq2}
bias_{u,k} = \sum_{i=1}^{N}(bias_{u,i,k})
\end{equation}

\begin{equation}\label{joiner_eq_bias_modified}
    \begin{split}
	z^{t, u} & = \phi(U h^{enc}_{t}+ V[c_{u}; h^{pred}_{u}] + Dropout(B b_{u}) + b)
	\end{split}
\end{equation}

Alternatively, we can use value of 1 for $\alpha_{u,i}$ in Equation \eqref{biasingeq1} rather than computing it from AttModule  and achieve contextualization by using only BiasingModule. This results in a modification of the Joiner equation \eqref{joiner_eq} to equation   \eqref{joiner_eq_bias_modified_biasing_only}:

\begin{equation}\label{joiner_eq_bias_modified_biasing_only}
    \begin{split}
	z^{t, u} & = \phi(U h^{enc}_{t}+ Vh^{pred}_{u} + Dropout(B b_{u}) + b)
	\end{split}
\end{equation}
}

Equations~\ref{softamx_linear_eq} and~\ref{softamx_eq} remain unchanged.

\begin{figure*}[t]
  \centering
\includegraphics[width=\linewidth,height=8cm]{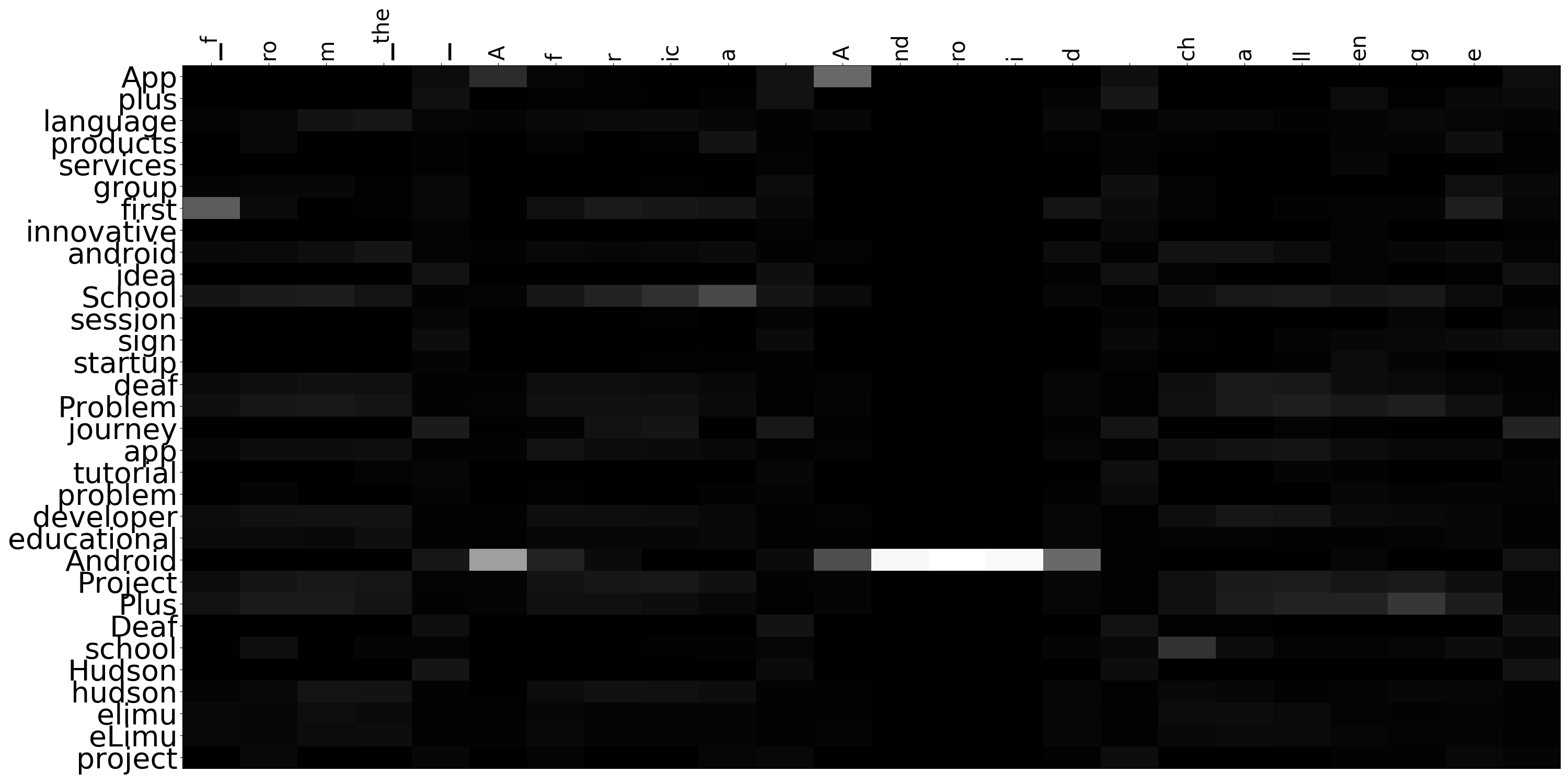}
 \caption{Visualizing attention weights, $\alpha_{u,n}$ from Equation~\eqref{att_softmax_eq}, for the example in Table \ref{tab:compare_examples}, row 1. The $x$-axis shows the target units from the hypothesis output by the Contextual RNN-T Model, and the $y$-axis shows the contextual metadata words ($w_{i}$). Darker colors represent values close to zero while brighter colors represent values closer to 1.}
  \label{fig:attention_vis1} 
\end{figure*}

\begin{table*}[thbp!]
 \vspace{-0.6em}
    \begin{tabular}{ | c | c | c | c|  }
     \hline
     Reference Snippet & Baseline Output &  Contextualization Output & Metadata Words (truncated)\\
    \hline
    \begin{tabular}
    {@{}c@{}} from the \textbf{Africa Android} \\  Challenge  \end{tabular} & \begin{tabular}
    {@{}c@{}}from the \textbf{Africa} {\color{black} and red} \\ challenge \end{tabular}& from the \textbf{Africa Android} challenge & 
    \begin{tabular}
    {@{}c@{}}
    %Plus, app, tutorial \\  
    innovative, School, language, \\
    Plus, app, tutorial, \\  
    %first, session, Project \\  
    products, 
    educational,
    startup \\  Android, problem, ...\\ % android ,elimu, Deaf \\ developer, sign, idea \\  App ,project, hudson   journey 
    \end{tabular}  \\
    \hline
    \begin{tabular}
    {@{}c@{}}its very intuitive so when \\ you look  at \textbf{PyTorch} itself \end{tabular} &
    \begin{tabular}
    {@{}c@{}}its very intuitive so when \\ you look at  {\color{black} pie towards} itself \end{tabular} &  \begin{tabular}    {@{}c@{}} its very intuitive so when \\ you look at  \textbf{PyTorch} itself \end{tabular}
    & 
    \begin{tabular}
    {@{}c@{}}
     experiences, novel, PyTorch  \\
     %computer, processing, product  \\ %deployment, latest, Experiences  \\ 
     %path,    \\
     updates, Facebook, machine,  \\
     AI, language, research, ...  %\\ ai, facebook, Facebook  \\  new, scale, vision, production  \\  natural, translation    \\
    \end{tabular}  \\
        \hline
    \end{tabular}
  \label{tab:compare_examples}
   \captionsetup{font=scriptsize}
    \caption{\label{tab:compare_examples} Comparing outputs generated by the baseline and Contextual RNN-T models. Named entities are represented with bold font in these examples.}
  \vspace{-1em}
\end{table*}

\section{Experiments}
\label{sec:exp}

\subsection{Dataset}
\label{subsec:dataset}
The dataset used for our experiments was sampled from English videos shared publicly on Facebook. The data is de-identified by removing information such as the user who posted the video, and only use the content of the video and the text metadata for training and evaluation. 

We segment the data in duration of 10 seconds for training and evaluation. Metadata words are obtained from the title and description text of the video, if available, after doing simple text cleaning and filtering such as removing words with hyperlinks in them. If a word in the text is capitalized then we also add its lowered case version as metadata word. We do not preserve the ordering of words in metadata both in training and evaluations. Each segment of the same video shares the same list of contextual words. We use about 8k hours of data for training and about 170 hours for evaluation. We further divide the evaluation test set based on whether there is any word in the reference for that segment that also appears in the metadata words for that video. The segments for which there is at least one common word are referred to below as the \emph{CommonNonZero} set and the remainder as the \emph{CommonZero} set. We present results on these two evaluation sets.

\subsection{Model}
\label{subsec:model}
The architecture of the Contextual RNN-T model from Section~\ref{sec:rnnt_dc} used for the experiments in this paper is as follows. The Audio Encoder is a {\color{black}4-layer BLSTM with 604 dimensions. We use subsampling of 2 across the time dimension after the first and second BLSTM layers. The output of the last layer of the BLSTM is projected to 1024 dimensions. The Text Predictor is a 2-layer LSTM of 256 dimensions whose output is also projected to 1024 dimensions}. We used a token set consisting of 200 sentence pieces, trained using the sentence piece library \cite{sentence_piece}. The Embedding Extractor is a 2-layer BLSTM of size 100. The Attention Module has the following parameters: 1) Convolution ($Q$) with 2 out channels and kernel of size 1, 2) Attention is computed over 64 dimensions with $A$ being of size $1024*64$, $B$ of size $200*64$ and $C$ of size $2*64$ (in Equations \eqref{att_conv_eq}, \eqref{att_weight_eq} and \eqref{att_softmax_eq}). {\color{black}The output of the Biasing Module, when used, is also linearly projected to 1024 dimensions (Equation \eqref{joiner_eq_bias_modified}}). The baseline model (Figure \ref{fig:rnnt_fig}) does not use Embedding Extractor, Attention Module {\color{black} or Biasing Module}. All components of both the baseline and Contextual RNN-T models are trained from scratch.

The input to the network consists of globally normalized 80-dimensonal log Mel-filterbank features, extracted with 25ms FFT windows and 10ms frame shifts. Sentence piece encoding of each word, $w_{i}$, in the metadata is appended with a special sentence piece unit. We use the Adam optimizer~\cite{kingma2014adam}, with a learning rate of 0.0002, and SpecAugment~\cite{spec_aug} with policy LB during training. All models were trained for 80 epochs. A beam size of 10 was used during inference.  

\subsection{Impact on WER-NE and WER}

\label{sec:sec_impact_wer_eer}
We measure performance of our models using WER and WER-NE on the two test sets described in Section~\ref{subsec:dataset}. An in-house Entity tagger was used to tag named entities in transcripts and metadata. 

As seen in Table \ref{tab:tab_impact_nonzero}, the Contextual RNN-T model {\color{black} with AttModule} {\color{black}(Equation \eqref{joiner_eq_modified})} (row 2) improves on WER-NE by about {\color{black}13\%} relative compared to the baseline model (row 1) on the \emph{CommonNonZero} evaluation set. {\color{black}Introducing BiasingModule (as in Equation \eqref{joiner_eq_bias_modified}) improves WER-NE further by another 3\%} (row 4). {\color{black} The Contextual RNN-T model that uses only BiasingModule (Equation \eqref{joiner_eq_bias_modified_biasing_only}) improves on WER-NE by about 14\%  (row 3) relative compared to the baseline model (row 1). This shows that using either BiasingModule or AttModule gives comparable WER-NE improvement but combining them further improves WER-NE.} As shown in Table \ref{tab:tab_impact_zero}, both WER and WER-NE for the \emph{CommonZero} test set does not get significantly impacted by the Contextual RNN-T models when there is no intersection between the metadata words and the reference.

We also measure robustness of our system using precision and recall of the emission of context words in the model's hypotheses. A \emph{True Positive} occurs when a context word from the metadata of the video is correctly output by the model as compared to the reference. A \emph{False Positive} occurs if the model outputs a context word but it does not appear in the reference.  We show aggregated precision and recall over both test sets for triggering of the context words in Table  \ref{tab:true_false_rate}. We see an improvement in recall of {\color{black} 8.3\% and degradation in precision by 1.3\% relative} for the {\color{black}best contextual} model compared to the baseline.

\vskip 0.06in

\begin{table}[thbp!]
\centering
 \vspace{-0.9em}
    \begin{tabular}{ | c | c |  c | c | }
    \hline
       Model  & WER  & WER-NE  \\ 

    \hline
    Baseline & 16.04 & 24.69   \\
    \hline %f190140565
    with AttModule & 15.45 & 21.41 \\ 
             \hline
with  BiasingModule & 15.51 & 21.14 \\
    \hline
    with AttModule and BiasingModule & 15.37 & 20.66 \\
  
    \hline %f190140926
    \end{tabular}
  \label{tab:tab_impact_nonzero}
  \captionsetup{font=scriptsize}
    \caption{\label{tab:tab_impact_nonzero} {\it WER and WER-NE results on CommonNonZero test set} }
  \vspace{-0.9em}
\end{table}

\begin{table}[thbp!]
\centering
 \vspace{-0.9em}
    \begin{tabular}{ | c | c |  c | c | }
    \hline
       Model  & WER  & WER-NE  \\ 

    \hline
    Baseline & 23.07  & 29.18  \\ %f190140465
    \hline
   with AttModule & 22.95 & 29.71 \\
  \hline
  with BiasingModule & 23.13 & 29.89 \\
 \hline
  with AttModule and BiasingModule & 22.89 & 29.93 \\

    \hline %f190140761
  %  \hline
    \end{tabular}
  \label{tab:tab_impact_zero}
  \captionsetup{font=scriptsize}
    \caption{\label{tab:tab_impact_zero} {\it WER and WER-NE results on CommonZero test set} }
  \vspace{-0.9em}
\end{table}

\begin{table}[thbp!]
\centering
 \vspace{-0.9em}
    \begin{tabular}{ | c | c |  c| }
         \hline
    Model & Precision & Recall \\   \hline
    Baseline &  0.934 & 0.844 \\   \hline
    with AttModule & 0.920 &  0.898 \\    \hline
  with BiasingModule & 0.918 & 0.902 \\ 
    \hline
   with AttModule and BiasingModule & 0.922 &  0.914 \\ 
\hline
    \end{tabular}
  \label{tab:context_eff}
  \captionsetup{font=scriptsize}
   \caption{\label{tab:true_false_rate} {\it Precision and Recall for context words across both test sets} }
  \vspace{-0.9em}
\end{table}

\section{Analysis}
\label{sec:analysis}
To understand better what the Contextual RNN-T model is doing, we visualize attention values for a few test segments where it correctly recognizes named entities that the baseline model makes errors on. These examples are shown in Table \ref{tab:compare_examples}.

For the example shown in row 1 of Table \ref{tab:compare_examples}, both the Contextual and baseline models are able to recognize common entities such as \textbf{Africa}. However, the baseline model has difficulties in recognizing entities that are not frequent in training data set, such as \textbf{Android} and \textbf{PyTorch}. Since \textbf{Android} appears in the metadata, the Contextual RNN-T model is able to attend to it and transcribe it correctly. This can be seen in the visualization of attention given to the words in the metadata at each output target unit ($u$) of the Contextual RNN-T Model in Figure \ref{fig:attention_vis1}. 

\section{Conclusion}
\label{sec:conclusion}
%{\color{black}In this work, we show that contextual metadata text, even if it is noisy, can be used to improve recognition of named entities. Some directions to explore further as future work could be: i) Using contextual embeddings from other modalities such as images from video, ii) Using biasing module with deep contextual framework for other application such as domain adaption etc.
We show that contextual metadata text, even if it is noisy, can be used to improve recognition of named entities for a challenging open domain ASR task such as social media videos within the framework of an E2E RNN-T ASR model. Some future explorations could be: i) Using contextual embeddings from other modalities such as images from video, ii) Using semantic embeddings to represent the metadata.

\section{Acknowledgement}
\label{sec:acknowledgement}

The authors would like to thank Anuroop Sriram and Duc Le for many helpful discussions and suggestions.

%Note from MJ: introducing ack section back which i removed for space preserving in final version submitted for conference.

\bibliographystyle{IEEEtran}

\bibliography{template}

\begin{thebibliography}{10}
\providecommand{\url}[1]{#1}
\csname url@samestyle\endcsname
\providecommand{\newblock}{\relax}
\providecommand{\bibinfo}[2]{#2}
\providecommand{\BIBentrySTDinterwordspacing}{\spaceskip=0pt\relax}
\providecommand{\BIBentryALTinterwordstretchfactor}{4}
\providecommand{\BIBentryALTinterwordspacing}{\spaceskip=\fontdimen2\font plus
\BIBentryALTinterwordstretchfactor\fontdimen3\font minus
  \fontdimen4\font\relax}
\providecommand{\BIBforeignlanguage}[2]{{%
\expandafter\ifx\csname l@#1\endcsname\relax
\typeout{** WARNING: IEEEtran.bst: No hyphenation pattern has been}%
\typeout{** loaded for the language `#1'. Using the pattern for}%
\typeout{** the default language instead.}%
\else
\language=\csname l@#1\endcsname
\fi
#2}}
\providecommand{\BIBdecl}{\relax}
\BIBdecl

\bibitem{hybrid-ctc}
H.~A. Bourlard and N.~Morgan, \emph{Connectionist speech recognition: a hybrid
  approach}.\hskip 1em plus 0.5em minus 0.4em\relax Springer Science \&
  Business Media, 2012, vol. 247.

\bibitem{rnnt-graves}
A.~Graves, ``Sequence transduction with recurrent neural networks,''
  \emph{arXiv preprint arXiv:1211.3711}, 2012.

\bibitem{las}
W.~Chan, N.~Jaitly, Q.~Le, and O.~Vinyals, ``Listen, attend and spell: A neural
  network for large vocabulary conversational speech recognition,'' in
  \emph{2016 IEEE International Conference on Acoustics, Speech and Signal
  Processing (ICASSP)}.\hskip 1em plus 0.5em minus 0.4em\relax IEEE, 2016, pp.
  4960--4964.

\bibitem{loc_attn}
J.~K. Chorowski, D.~Bahdanau, D.~Serdyuk, K.~Cho, and Y.~Bengio,
  ``Attention-based models for speech recognition,'' in \emph{Advances in
  neural information processing systems}, 2015, pp. 577--585.

\bibitem{hinton2012deep}
G.~Hinton, L.~Deng, D.~Yu, G.~Dahl, A.-r. Mohamed, N.~Jaitly, A.~Senior,
  V.~Vanhoucke, P.~Nguyen, B.~Kingsbury \emph{et~al.}, ``Deep neural networks
  for acoustic modeling in speech recognition,'' \emph{IEEE Signal processing
  magazine}, vol.~29, 2012.

\bibitem{mohri2002weighted}
M.~Mohri, F.~Pereira, and M.~Riley, ``{Weighted finite-state transducers in
  speech recognition},'' \emph{Computer Speech \& Language}, vol.~16, no.~1,
  pp. 69--88, 2002.

\bibitem{clas}
G.~Pundak, T.~N. Sainath, R.~Prabhavalkar, A.~Kannan, and D.~Zhao, ``Deep
  context: end-to-end contextual speech recognition,'' in \emph{2018 IEEE
  Spoken Language Technology Workshop (SLT)}.\hskip 1em plus 0.5em minus
  0.4em\relax IEEE, 2018, pp. 418--425.

\bibitem{dc_tom}
Z.~Chen, M.~Jain, Y.~Wang, M.~L. Seltzer, and C.~Fuegen, ``Joint grapheme and
  phoneme embeddings for contextual end-to-end asr,'' in \emph{INTERSPEECH},
  2019.

\bibitem{dc_phoebe}
A.~{Bruguier}, R.~{Prabhavalkar}, G.~{Pundak}, and T.~N. {Sainath}, ``Phoebe:
  Pronunciation-aware contextualization for end-to-end speech recognition,''
  2019.

\bibitem{shallowe2e}
D.~Zhao, T.~N. Sainath, D.~Rybach, D.~Bhatia, B.~Li, and R.~Pang,
  ``Shallow-fusion end-to-end contextual biasing,'' \emph{Submitted to
  Interspeech}, vol. 2019, 2019.

\bibitem{shallowfusion2018}
I.~Williams, A.~Kannan, P.~S. Aleksic, D.~Rybach, and T.~N. Sainath,
  ``Contextual speech recognition in end-to-end neural network systems using
  beam search.'' in \emph{Interspeech}, 2018, pp. 2227--2231.

\bibitem{streaming_rnnt}
Y.~He, T.~N. Sainath, R.~Prabhavalkar, I.~McGraw, R.~Alvarez, D.~Zhao,
  D.~Rybach, A.~Kannan, Y.~Wu, R.~Pang \emph{et~al.}, ``Streaming end-to-end
  speech recognition for mobile devices,'' in \emph{ICASSP 2019-2019 IEEE
  International Conference on Acoustics, Speech and Signal Processing
  (ICASSP)}.\hskip 1em plus 0.5em minus 0.4em\relax IEEE, 2019, pp. 6381--6385.

\bibitem{keyword-rnnt}
Y.~He, R.~Prabhavalkar, K.~Rao, W.~Li, A.~Bakhtin, and I.~McGraw, ``Streaming
  small-footprint keyword spotting using sequence-to-sequence models,'' in
  \emph{2017 IEEE Automatic Speech Recognition and Understanding Workshop
  (ASRU)}.\hskip 1em plus 0.5em minus 0.4em\relax IEEE, 2017, pp. 474--481.

\bibitem{darong-paper}
D.-R. Liu, C.~Liu, F.~Zhang, G.~Synnaeve, Y.~Saraf, and G.~Zweig,
  ``Contextualizing asr lattice rescoring with hybrid pointer network language
  model,'' in \emph{Submitted ro Proc. Interspeech}, 2020.

\bibitem{relu}
V.~Nair and G.~E. Hinton, ``Rectified linear units improve restricted boltzmann
  machines,'' in \emph{Proceedings of the 27th international conference on
  machine learning (ICML-10)}, 2010, pp. 807--814.

\bibitem{graves2006connectionist}
A.~Graves, S.~Fern{\'a}ndez, F.~Gomez, and J.~Schmidhuber, ``Connectionist
  temporal classification: labelling unsegmented sequence data with recurrent
  neural networks,'' in \emph{Proceedings of the 23rd international conference
  on Machine learning}.\hskip 1em plus 0.5em minus 0.4em\relax ACM, 2006, pp.
  369--376.

\bibitem{sentence_piece}
T.~Kudo and J.~Richardson, ``Sentencepiece: A simple and language independent
  subword tokenizer and detokenizer for neural text processing,'' \emph{arXiv
  preprint arXiv:1808.06226}, 2018.

\bibitem{kingma2014adam}
D.~P. Kingma and J.~Ba, ``Adam: A method for stochastic optimization,''
  \emph{arXiv preprint arXiv:1412.6980}, 2014.

\bibitem{spec_aug}
D.~S. Park, W.~Chan, Y.~Zhang, C.-C. Chiu, B.~Zoph, E.~D. Cubuk, and Q.~V. Le,
  ``Specaugment: A simple data augmentation method for automatic speech
  recognition,'' \emph{arXiv preprint arXiv:1904.08779}, 2019.

\end{thebibliography}

% \begin{thebibliography}{9}
% \bibitem[1]{Davis80-COP}
%   S.\ B.\ Davis and P.\ Mermelstein,
%   ``Comparison of parametric representation for monosyllabic word recognition in continuously spoken sentences,''
%   \textit{IEEE Transactions on Acoustics, Speech and Signal Processing}, vol.~28, no.~4, pp.~357--366, 1980.
% \bibitem[2]{Rabiner89-ATO}
%   L.\ R.\ Rabiner,
%   ``A tutorial on hidden Markov models and selected applications in speech recognition,''
%   \textit{Proceedings of the IEEE}, vol.~77, no.~2, pp.~257-286, 1989.
% \bibitem[3]{Hastie09-TEO}
%   T.\ Hastie, R.\ Tibshirani, and J.\ Friedman,
%   \textit{The Elements of Statistical Learning -- Data Mining, Inference, and Prediction}.
%   New York: Springer, 2009.
% \bibitem[4]{YourName17-XXX}
%   F.\ Lastname1, F.\ Lastname2, and F.\ Lastname3,
%   ``Title of your INTERSPEECH 2020 publication,''
%   in \textit{Interspeech 2020 -- 20\textsuperscript{th} Annual Conference of the International Speech Communication Association, September 15-19, Graz, Austria, Proceedings, Proceedings}, 2020, pp.~100--104.
% \end{thebibliography}

\end{document}